# The collaboration behaviors of scientists in Italy: a field level analysis[1]

Giovanni Abramo

*National Research Council of Italy (IASI-CNR) and Laboratory for Studies of Research and Technology Transfer at University of Rome "Tor Vergata" – Italy*

ADDRESS: Dipartimento di Ingegneria dell'Impresa, Università degli Studi di Roma "Tor Vergata", Via del Politecnico 1, 00133 Roma – ITALY
tel. and fax +39 06 72597362, giovanni.abramo@uniroma2.it

Ciriaco Andrea D'Angelo

*Laboratory for Studies of Research and Technology Transfer at University of Rome "Tor Vergata" – Italy*

ADDRESS: Dipartimento di Ingegneria dell'Impresa, Università degli Studi di Roma "Tor Vergata", Via del Politecnico 1, 00133 Roma – ITALY
tel. and fax +39 06 72597362, dangelo@dii.uniroma2.it

Gianluca Murgia

*University of Siena – Italy*

ADDRESS: Dipartimento di Ingegneria dell'Informazione e Science Matematiche, Università degli Studi di Siena, Via Roma 56, 53100 Siena – ITALY
tel. and fax +39 0577 1916386, murgia@dii.unisi.it

# The collaboration behaviors of scientists in Italy: a field level analysis

**Abstract**

The analysis of research collaboration by field is traditionally conducted beginning with the classification of the publications from the context of interest. In this work we propose an alternative approach based on the classification of the authors by field. The proposed method is more precise if the intended use is to provide a benchmark for the evaluation of individual propensity to collaborate. In the current study we apply the new methodology to all Italian university researchers in the hard sciences, measuring the propensity to collaborate for the various fields: in general, and specifically with intramural colleagues, extramural domestic and extramural foreign organizations. Using a simulation, we show that the results present substantial differences from those obtained through application of traditional approaches.

# 1. Introduction

Over the past several decades there has been a remarkable growth in collaboration for the purposes of scientific research. The reality has often been confirmed by analysis of co-authorship (Melin and Persson, 1996), which indicates that the share of single-authored publications is constantly on the decline (Abt, 2007; Uddin et al., 2012).

However the actual modalities of collaboration (intramural/extramural, domestic/international, intradisciplinary/interdisciplinary) can vary in intensity on the basis of contextual factors, first of all with the research discipline involved (Gazni et al., 2012; Yoshikane and Kageura, 2004). For example, the so-called "big science" disciplines show publications with a much higher number of authors than those typically seen in other disciplines, due to factors such as the cost of equipment and necessity of large sample sizes, and also the manner of assigning publication authorship (Cronin, 2001; Glanzel and Schubert, 2004).

Even within a single discipline there can be notable heterogeneity in the forms of activating collaboration, due to the different specializations involved (Piette and Ross, 1992), as well as the different propensities to collaborate of the individual scientists (Newman, 2001; Moody, 2004). Knowledge of the different forms of activating collaborations across fields and disciplines permits investigation of the mechanisms at the very base of collaboration and definition of the most suitable policies for its management, potentially for increased research productivity (Wagner and Leydesdorff, 2005).

In this article we will investigate the collaboration behaviors of scientists from different disciplines. In the literature, such studies are generally based on the initial step of classifying the publications by subject category. Our approach instead begins with the disciplinary classification of the scientists. An unusual characteristic of the Italian academic system makes this approach possible. In what seems to be a unique situation, the Italian Ministry of Universities and Research (MIUR) maintains a database[2] of all national academics, in which each individual is classified in one and only one Scientific Disciplinary Sector (SDS). There are 370 such fields[3], in turn grouped under 14 University Disciplinary Areas (UDAs). By associating each publication with its respective authors, it is possible to then compare the propensity to collaborate in different forms, for scientists in different fields and disciplines. By applying the traditional methodology based on the classification of publications for the same population, we are then able to measure the difference in the results derived from the two methods.

In a preceding work, Abramo et al. (2012) applied the same investigative approach to interdisciplinary collaboration in public research organizations, to test for returns to scope of research fields. By incidence, this type of examination would have been impossible using traditional methods based on classification of the publications. In this work we again wish to use analysis of co-authorships to examine other aspects of research collaborations, in particular to detect any potential difference among scientists from different fields in their propensities to collaborate in general, and in intramural, extramural domestic and extramural international collaborations.

---

[2] http://cercauniversita.cineca.it/php5/docenti/cerca.php, last access January 28, 2013.
[3] The complete list is accessible at http://attiministeriali.miur.it/UserFiles/115.htm, last accessed January 28, 2013.

In the following section we review the relevant literature, and in Section 3 describe the methodology and field of observation. In sections 4 and 5 we first illustrate the results from application of the "author-classification" methodology, then measure the shifts in results compared to those from the traditional method. In the final section we propose further avenues of investigation and provide several policy recommendations. Observing collaboration patterns by authors' classification rather than by publications' classification data leads to more precise measures of individual propensity to collaborate in the various forms, because of the very skewed productivity distribution, with few scientists determining most of the observations. Further, the choice of focusing on individual scientists also permits an understanding of the links that exist between the propensities for the various forms of collaboration.

## 2. Scientific collaboration and its determinants

One of the critical factors for development of productive scientific collaboration is the possibility of communicating in an effective, informal and flexible manner, particularly in the early phases when it is necessary to increase familiarity and create a climate of trust among collaborators (Traore and Landry, 1997). Thus it is no surprise that a large part of collaborations are activated through face-to-face encounters, such as discussions in the workplace, at conferences or in purpose-planned start-up meetings, where communication among participants is freer and less conditioned by means of communication (Laudel, 2001; Wagner and Leydesdorff, 2005). Face-to-face encounters can also help to ease problems in coordination during implementation phases, avoiding problems of "freeriding" and conflict among partners, especially in long-distance collaboration where monitoring is more difficult (Hinds and Bailey, 2003). The importance of face-to-face encounters could be at the root of the phenomena in which propensity to collaborate diminishes with geographic distance between the home organizations of the scientists (Hoekman et al., 2010; Abramo et al., 2009; Larivière et al., 2006). This factor would also explain the different forms of co-authorship for scientists that belong to different sizes of universities (Katz, 2000): those from large universities tend to collaborate primarily with colleagues from the same university or from foreign organizations, while those from smaller universities tend to work with colleagues belonging to other domestic universities, given the limited number of their own intramural colleagues and their lesser "relational" and financial resources.

The increase in scientific collaborations in recent years, especially at the international level, is most certainly related to the general reduction in travel costs (Hoekman et al., 2010). However, the single most important factor in the notable increase in extramural scientific collaborations is the diffusion of inexpensive new communications technologies, which greatly reduce the qualitative divide between distant and face-to-face communication, even though still not completely eliminating the differences (Cairncross, 1997; Olsen and Olsen, 2000).

There are also factors acting as disincentives to scientific collaboration, particularly at the international level. The legal and regulatory frameworks concerning project management and research results can lead to scientists avoiding collaboration, especially in the applied sciences (Jeong et al., 2011). Language and cultural barriers represent further potential obstacles (Schubert and Glanzel, 2006; Zitt et al., 2000). Even within a single country there can be cultural barriers between organizations and disciplines, to the point that some scientists collaborate only within their so-called "invisible college", composed of colleagues originating from the same institution, thus sharing similar skills, scientific approaches and professional relations (Crane, 1969; Bozeman and Corley, 2004).

The choice among forms for collaboration can also be influenced by incentives and considerations concerning the organizational research system. For example, a scientist could be encouraged to privilege intramural collaborations because these will favor creation of team spirit within the workplace (Acedo et al., 2006). Opting for intramural collaborations could also have positive effects for the scientists' career, especially when access to higher positions is either in whole or in part managed at the local level. In other cases, mechanisms for career advancement that reward external collaboration could instead encourage collaborations with colleagues from other domestic and foreign organizations (Wagner and Leydesdorff, 2005).

The forms and sources of research financing can also influence the types of collaboration chosen: financing from regional and national agencies and enterprises can encourage internal or local-level collaborations (Jeong et al., 2011), while supra-national financing, and in certain cases the incentive systems internal to individual organizations, can favor collaborations at the international level (Gossart and Ozman, 2009; Hoekman et al., 2010). Incentive systems for international scientific collaboration accomplish political ends (Banda, 2000) and also bring about more productive research at the practical level, especially in some of the experimental disciplines such as physics and the life sciences, which require costly equipment and large samples for statistical significance, obtainable only through formation of international institutional networks (Laband and Tollison, 2000; Gazni et al., 2012). In research on specifically national themes of interest, the most frequent collaborations involve scientists belonging to organizations from the particularly country, thus achieving greater knowledge transfer and benefits in the national economy (Gossart and Ozman, 2009).

The choice of collaboration form does not depend only on the characteristics of the discipline, but also on the interdisciplinary nature of each individual project (Persson et al., 1997). If the necessary competencies are not present within an organization, interdisciplinary research requires collaboration agreements with third parties and thus leads to shared publication of the research results (Katz and Martin, 1997). In the same way, scientists may seek out collaborations with prestigious colleagues from other organizations if they find that their own universities do not offer potential partners of similar outstanding reputation (Jeong et al., 2011). This tendency explains how more prestigious universities have a greater number of collaborations compared to others (Piette and Ross, 1992) and why, in a similar way, more advanced nations have a central role in international collaboration networks (Luukkonen et al., 1992).

The choice of co-authorship, especially involving individuals of different competencies, cultures and previous experiences, can increase group creativity and satisfy the requisites of complexity and interdisciplinarity that are particularly expected for certain research themes (He et al., 2009). As proof, various studies have shown how co-authored publications, especially at the international level, achieve above-average impact measured both in terms of journal importance (Bordons et al., 1996) and citations received (Hoekman et al., 2010; Jones et al., 2008). He et al. (2009) further show how international collaborations primarily influence the quality of a scientist's publications, while "domestic" collaborations influence publication intensity. Glanzel and Schubert (2004) show that the researcher's choice of intramural collaboration tends to lead to the situation of "ghost authorship", meaning lack of recognition among the coauthors, more so than occurs in choices for inter-university or international collaborations.

The whole of the research cited shows how the choices of collaborating with different colleagues from the same university, from organizations in the home nation or abroad, are all linked to different factors, of financial, social and other character, with weights that vary within each discipline. The development of adequate instruments for analyzing the forms of collaboration that are currently prevalent among the scientists belonging to each discipline thus represents an indispensable step for the definition of adequate policies to manage the phenomenon.

## 3. Methodology, dataset and indicators

*3.1 Methodology*

Research collaborations are generally studied by defining the form of collaboration (intramural or extramural, intra-disciplinary or interdisciplinary, public-private, domestic or international, etc.), the context of the analysis, such as a discipline or universities, and the instrument of analysis, which is often the co-authorships of the publications. All the publications that can be referred to the specified context are then classified on the basis of the means of collaboration that are the object of study. For example Gazni et al. (2012) classify the publications referable to a discipline as "international" on the basis of the presence or absence of an author belonging to a foreign organization. To evaluate the incidence of international collaboration within the discipline they calculate the ratio of the number of publications classified as "international" to the total of publications in the discipline. This is the approach that underlies all the principle indicators of co-authorship developed in the literature, beginning with "Degree of Collaboration" by Subramanyam (1983), then "Collaborative Index" by Lawani (1986), "Collaborative Coefficient" by Ajiferuke et al. (1988), and finally Egghe's (1991) "Revised Collaborative Coefficient".

An alternative approach for study of co-authorship consists of using a base unit of analysis to aggregate the data concerning publications. The use of the single scientist as the base analytical unit permits evaluation of the propensity of the scientists, referred to the particular context of analysis, to collaborate in the form that is the object of study. Returning to the phenomenon examined by Gazni et al. (2012), the use of single scientists as the base analytical unit would permit evaluation of the propensity to international collaboration for the scientists that belong to a discipline. To our knowledge, this type of approach was used by Martin-Sempere et al. (2002) and Abramo et al. (2011). The latter measured the propensity to international collaboration for Italian academics, by discipline: for each academic, they calculated the ratio of number of publications co-authored with colleagues belonging to a foreign organization to the total number of his or her publications. Martin-Sempere et al., although limiting their study to 93 geologists from universities in Spain, calculated the "degree of collaboration", equal to the ratio of the number of coauthored publications to the total number of publications of a scientist, and the "degree of national collaboration", equal to the ratio of the number of publications coauthored with colleagues belonging to at least one national organization, to the total number of the scientist's publications.

Our current work analyzes the following forms of collaboration: intramural, extramural domestic and extramural international. The analysis refers to all Italian university professors in the hard sciences and some fields of the social sciences, where publications indexed by bibliometric databases represent a good proxy of overall research output (Moed, 2005). This is not the case for the arts and humanities, where the coverage of bibliometric databases are too limited. The instrument of analysis is the co-authorship of scientific publications over the period 2006-2010 as indexed on the Web of Science (WoS). We first calculate the different propensities to collaborate for the individual scientists, then define the distribution of these propensities among academics belonging to each discipline, and then for the fields within each discipline, in order to show the differences in forms of collaboration selected in the various disciplines. This method also allows to test the correlation between the different propensities to collaborate, for the academics belonging to the various disciplines. Finally, we analyze the differences between our indicators of propensity and those generally used in the literature, measuring the same forms of collaboration but beginning from the classification of the publications in fields instead of the classification of the authors.

*3.2 Data sources and field of observation*

The dataset of Italian professors used in our analysis was extracted from the Ministry of Universities and Research database, which we have described above. Next, the dataset of these individuals' publications is extracted from the Italian Observatory of Public Research (ORP), a database developed and maintained by the authors and derived under license from the WoS. Beginning from the raw data of 2006-2010 Italian publications in WoS, and applying a complex algorithm for disambiguation of the true identity of the authors and their institutional affiliations (for details see D'Angelo et al., 2011), each publication[4] is attributed to the university scientist or scientists (full, associate and assistant professors) that produced it, with a harmonic average of precision and recall (F-measure) equal to 96 (error of 4%).

For each publication, the bibliometric dataset thus provides:
- the complete list of all coauthors;
- the complete list of all their addresses;
- a sub-list of only the academic authors, with their SDS/UDA and university affiliations.

---

[4] We exclude those document types that cannot be strictly considered as true research products, such as editorial material, conference abstracts, replies to letters, etc.

Our dataset permits unequivocal identification of each academic with their home university, although this operation is not possible for non-academic authors of the publications. It is also not possible to associate the academics with any organizations other than their own universities, although the literature shows (Katz and Martin, 1997) that in some cases authors indicate more than one institutional address, due to some form of multiple engagement or change in employment. This can actually lead to certain problems, such classifying publications as being produced under international co-authorship when the presence of a foreign organization in the byline is actually due to a single academic belonging to multiple organizations (Glanzel, 2001). Further, our dataset permits unequivocal assignment of every academic to their SDS, and thus to the UDA to which they belong, while the same operation is not possible for nonacademic authors of the publications. For these reasons, the analysis is conducted only for university researchers.

Table 1 presents the statistics for the population of Italian academics belonging to the 11 UDAs analyzed and their respective publications. To render the bibliometric analysis still more robust, the field of observation is limited to those SDSs (200 in all) where at least 50% of academics produce at least one publication in the 2006-2010 period.

The number of academics belonging to each UDA varies substantially, as does the percentage of "productive" academics (producing at least one publication indexed under the WoS in the period 2006-2010) and the percentage of "collaborative" academics (at least one publication in co-authorship with other scientists in the same period).

*Table 1: Main characteristics of the population of academics analyzed*

| UDA | Publications | Research staff | | |
| --- | --- | --- | --- | --- |
| | | Total | Productive | Collaborative |
| Medicine (MED) | 63,018 | 12,433 | 10,184 (81.9%) | 10,174 (81.8%) |
| Industrial and information engineering (IIE) | 37,283 | 5,644 | 4,846 (85.9%) | 4,822 (85.4%) |
| Biology (BIO) | 31,277 | 5,855 | 5,244 (89.6%) | 5,237 (89.4%) |
| Chemistry (CHE) | 25,687 | 3,610 | 3,384 (93.7%) | 3,379 (93.6%) |
| Physics (PHY) | 23,702 | 2,873 | 2,602 (90.6%) | 2,575 (89.6%) |
| Mathematics and computer sciences (MAT) | 16,131 | 3,607 | 2,905 (80.5%) | 2,809 (77.9%) |
| Agricultural and veterinary sciences (AVS) | 11,767 | 3,183 | 2,720 (85.5%) | 2,714 (85.3%) |
| Civil engineering (CEN) | 5,371 | 1,747 | 1,230 (70.4%) | 1,209 (69.2%) |
| Earth Sciences (EAR) | 5,284 | 1,423 | 1,181 (83.0%) | 1,173 (82.4%) |
| Economics and statistics (ECS) | 3,579 | 1,949 | 1,200 (61.6%) | 1,100 (56.4%) |
| Pedagogy and psychology (PPS) | 3,345 | 1,055 | 715 (67.8%) | 705 (66.8%) |
| Total | 197,460* | 43,379 | 36,211 (83.5%) | 35,897 (82.8%) |

* The total is less than the sum of those from the different UDAs, since 28,984 publications are in co-authorship by academics from different UDAs.

The productive academics are more than 80% of the overall total, but this value falls to just over 60% for academics in the Economics and statistics UDA and peaks at over 90% for Chemistry. Among other reasons, this variation is due to the fact that scientists belonging to some UDAs tend to publish research not only in journals censused by the WoS, but also in other journals, conference papers and books that are sometimes only of national interest (Larivière et al., 2006).

The collaborative academics are over 90% of the total of productive ones. The greatest difference between percentages of productive and collaborative scientists is in Economics and statistics, while this difference reaches the lowest level for the Medicine UDA. In this case, the variation is due to the fact that some scientists belonging to certain UDAs tend more to produce all their publications alone, although this phenomenon has been decreasing in recent years (Uddin et al., 2012).

*3.3 Indicators and methods*

Beginning from the individual researcher of known scientific field, we will compare the average propensity to collaborate in the different fields for each of four forms: in general; intramural; and extramural with researchers from domestic and foreign organizations. The first form of collaboration, the propensity to collaborate in general, represents a superset of the other forms.

We construct an "author-publication" matrix of dimensions *m* x *n* (36,211 x 197,460), equal to the total number of productive academics and the overall total of their overall publications. We then associate each academic with his or her publications (*p*) over the period. Since for each publication we know the number of authors and the numbers of Italian and foreign organizations, for each academic we can calculate the number of publications resulting from collaborations (*cp*), the number of publications resulting from collaborations with other academics belonging to the same university (intramural - *cip*), the number of publications from collaborations with scientists belonging to other domestic organizations (extramural domestic - *cedp*), and the number of publications with scientists belonging to foreign organizations (extramural international - *cefp*). From these values we can construct the indicators for the relative individual propensities to collaborate, from which we can then also obtain the average propensities per field and discipline:

- Propensity to collaborate $C = \frac{cp}{p}$
- Propensity to collaborate intramurally $CI = \frac{cip}{p}$
- Propensity to collaborate extramurally at the domestic level $CED = \frac{cedp}{p}$
- Propensity to collaborate extramurally at the international level $CEF = \frac{cefp}{p}$

Each of the four indicators varies between zero if, in the observed period, the scientist under observation did not produce any publications resulting from the form of collaboration analyzed, and 1 if the scientist produced all his or her publications through that form of collaboration[5].

**4. Results and discussion**

---

[5] Single authored papers with more than one affiliation are not considered as collaborations. A publication with more than two authors could present different forms of collaboration, for example intramural and extramural domestic. In this case it is counted in calculating propensity for both intramural and extramural domestic collaboration.

The calculation of *C*, *CI*, *CED* and *CEF* permits the analysis of the different forms of co-authorship and the characterization of the different UDAs and their individual SDSs on the basis of the values of propensity registered for their member academics. These steps of our analysis are reported in sections 4.1 and 4.2. In section 4.3 we then examine the correlation between the four indicators.

*4.1 Propensity to collaborate in different forms, in the various disciplines*

The academics belonging to the various analyzed UDAs show different propensities to collaboration in general and to intramural, extramural domestic and extramural international collaborations in specific. To analyze these differences, we present a table for each form of collaboration, showing per UDA: i) the values of average propensity to collaborate among the academics belonging to the UDA; ii) the percentage of academics with nil propensity; iii) percentage of academics with maximum (100%) propensity. To further validate the differences between the propensities registered for the academics belong to each discipline, we also apply the Kruskal-Wallis test (Kruskal and Wallis, 1952) for all the UDAs, and the Mann-Whitney U test (Mann and Whitney, 1947) for the various pairs of UDAs. These non-parametric tests permit verification of whether the level of academics' propensity to collaborate is more or less in one UDA or another. We conduct this analysis through the kruskal.test and the wilcox.test functions, and the results (accessible at Supplemental Material – S1) show highly significance for almost all the comparisons conducted[6]. Findings permit clustering of the UDAs on the basis of their different propensities to collaborate.

Table 2 shows the values of propensity to collaborate. These generally appear extremely high, in line with various other studies conducted using different approaches, all showing that the percentage of co-authored publications within the "bibliometric" disciplines is now near 90% (Abt, 2007; Gazni et al., 2012). The data in Table 2 show limited differences between many UDAs in the propensity to collaborate, although the results from the Mann-Whitney U test are often highly significant. In particular, the average propensity to collaborate reaches maximum values of near 100% in Medicine, Agricultural and veterinary sciences, Biology and Chemistry. These results substantially confirm those obtained by Haiqi and Hong (1997) and by Gazni et al. (2012), although the latter authors show a tendency towards increasing homogeneity between the different disciplines. The results obtained by Gazni and Didegah (2011) are also generally in line with what is observed in our research, apart from the disciplines falling under the Agricultural and veterinary sciences UDA, which in their research show lower values than Mathematics and computer science UDA. Instead, in our research this latter UDA, together with Economics and statistics, shows the least propensity to collaborate. This result can be explained above all in considering that the research conducted in these disciplines is primarily theory-based, rather than resource-based, thus generally requiring less resort to collaboration (Gazni et al., 2012). The fact that roughly 30% of the academics belonging to these disciplines have produced at least one publication as sole author over the period could also be due to the need to demonstrate the personal capacity to produce valid publications, without assistance from other colleagues – a need which in some areas was historically quite strong.

---

[6] Cases of non-significance are very few and always for reason of very limited difference in value of propensity to collaborate.

We also note that Physics shows an intermediate value of propensity to collaborate compared to other UDAs, probably because of the impact of some specific fields within the discipline. We will analyze this aspect in greater detail in section 4.2, where we show the different propensities to collaborate for the various SDSs that compose this particular discipline.

*Table 2: Propensity to collaborate, per UDA (percentage values)*

| UDA | Mean C | % C = 0% | % C = 100% |
|---|---|---|---|
| Medicine (MED) | 99.4 | 0.1 | 94.8 |
| Chemistry (CHE) | 99.2 | 0.1 | 94.8 |
| Agricultural and veterinary sciences (AVS) | 99.1 | 0.2 | 95.7 |
| Biology (BIO) | 99.1 | 0.1 | 94.4 |
| Earth sciences (EAR) | 97.6 | 0.7 | 90.6 |
| Industrial and information engineering (IIE) | 97.1 | 0.5 | 85.5 |
| Pedagogy and psychology (PPS) | 96.7 | 1.4 | 89.8 |
| Physics (PHY) | 96.6 | 1.0 | 81.5 |
| Civil engineering (CEN) | 94.3 | 1.7 | 81.5 |
| Mathematics and computer sciences (MAT) | 89.1 | 3.3 | 68.6 |
| Economics and statistics (ECS) | 84.0 | 8.3 | 70.1 |
| Total | 97.2 | 0.9 | 89.0 |

The differences between the various UDAs appear much more pronounced for intramural collaborations. As we see in Table 3, there is a difference of almost 40 percentage points between the UDA registering the maximum value (Chemistry) and that with the minimum value (Economics and statistics). These results and the ones from the Mann-Whitney U test show how the propensity to collaborate with academics from the same university is again very high in the four UDAs of Medicine, Agricultural and veterinary sciences, Biology and Chemistry, which previously registered the maximum general propensity to collaborate. This result can be explained considering that the academics that belong to these disciplines often use laboratory facilities owned by their own university, which for financial reasons are also shared with other colleagues, thus favoring development of collaboration. The second highest propensity to this particular form of collaboration is for Industrial and information engineering. This result can be explained considering that it is another discipline where a large part of research requires resources such as laboratories, equipment and software, which are shared with colleagues at the same university and with whom collaboration then develops more easily. Further, various studies in the engineering sphere are the result of research projects commissioned by companies to academics who then tend to involve colleagues from the same university above those from others, so as to reduce costs of communication and administration and have greater impact on the territory. The values of propensity to intramural collaboration as calculated in this study are difficult to compare with those reported from other bibliometric studies, which only map research done exclusively through intramural collaboration. In this sense, the present study permits more extensive evaluation of the role of intramural collaborations[7], being able to detect and map such collaborations even when the related publications indicate other institutional addresses.

*Table 3: Propensity to intramural collaboration, per UDA (percentage values)*

| UDA | Mean CI | % CI = 0% | % CI = 100% |
|---|---|---|---|
| Chemistry (CHE) | 83.5 | 2.4 | 46.1 |
| Industrial and information engineering (IIE) | 82.2 | 3.9 | 46.9 |
| Agricultural and veterinary sciences (AVS) | 81.2 | 4.3 | 51.8 |
| Medicine (MED) | 81.1 | 3.6 | 45.9 |
| Biology (BIO) | 78.8 | 4.2 | 45.8 |
| Civil engineering (CEN) | 73.4 | 8.8 | 46.3 |
| Physics (PHY) | 66.7 | 8.7 | 29.2 |
| Earth sciences (EAR) | 62.0 | 11.4 | 31.1 |
| Pedagogy and psychology (PPS) | 59.6 | 18.2 | 35.8 |
| Mathematics and computer sciences (MAT) | 54.1 | 20.5 | 25.4 |
| Economics and statistics (ECS) | 44.0 | 36.0 | 26.7 |
| Total | 75.4 | 7.2 | 42.3 |

Extramural collaborations can be classified on the basis of the location of the extramural organization, as being within or foreign to the same nation. Table 4 presents the average values of propensity to extramural collaboration with scientists belonging to other domestic organizations. Again, the differences between the various disciplines are highly noticeable, with the difference of almost 50 percentage points between the UDAs registering maximum average propensity (Physics) and minimum average propensity (Industrial and information engineering).

*Table 4: Propensity to extramural collaboration at the national level, per UDA (percentage values)*

| UDA | Mean CED | % CED = 0% | % CED = 100% |
|---|---|---|---|
| Physics (PHY) | 72.5 | 5.6 | 24.9 |
| Medicine (MED) | 62.4 | 8.2 | 20.6 |
| Earth Sciences (EAR) | 58.6 | 13.1 | 23.3 |
| Biology (BIO) | 57.4 | 9.8 | 17.5 |
| Chemistry (CHE) | 49.8 | 8.3 | 9.3 |
| Pedagogy and Psychology (PPS) | 48.5 | 26.3 | 22.2 |
| Agricultural and veterinary sciences (AVS) | 47.1 | 17.3 | 13.9 |
| Economics and Statistics (ECS) | 38.0 | 38.9 | 19.1 |
| Mathematics and Computer Sciences (MAT) | 33.6 | 33.3 | 10.7 |
| Civil engineering (CEN) | 26.0 | 44.3 | 8.2 |
| Industrial and information engineering (IIE) | 24.8 | 33.1 | 5.3 |
| Total | 50.3 | 17.0 | 15.7 |

These results are in line with the previous analysis by Abramo et al. (2009a), concerning publications resulting from collaborations involving Italian universities in the period 2001-2003. The low value observed in the engineering disciplines, particularly when considered jointly with the result obtained for propensity to intramural collaboration, suggests how collaborations are very important for these disciplines but are primarily effected within the individual university. This could be due to the fact that a large part of research has a primarily local character and that the necessary resources are substantially available within each university. However, in Physics and the other "big science" disciplines, the resources necessary for research (e.g. equipment, numbers of observations, interdisciplinary competencies) also require the involvement of scientists that belong to organizations beyond a single university.

---

[7] In our study, some publications may not be classified as the result of intramural collaboration even when the authors are from the same university. This could happen only when, among the authors belonging to the same university, there is only one who holds a recognized faculty position while the other author(s) are not part of the faculty. We recall that our dataset does not indicate the home institution of authors that are non-faculty, and in such cases therefore cannot identify them as belonging to the same institution.

This explanation is further supported by the analysis of propensity to extramural collaboration with scientists belonging to foreign organizations, as synthesized and presented in Table 5.

*Table 5: Propensity to extramural collaboration at the international level, per UDA (percentage values)*

| UDA | Mean CEF | % CEF = 0% | % CEF = 100% |
|---|---|---|---|
| Physics (PHY) | 51.7 | 12.2 | 10.0 |
| Earth Sciences (EAR) | 32.8 | 34.3 | 7.9 |
| Pedagogy and Psychology (PPS) | 31.8 | 40.1 | 11.7 |
| Economics and Statistics (ECS) | 27.5 | 52.3 | 12.1 |
| Biology (BIO) | 27.1 | 28.9 | 3.6 |
| Mathematics and Computer Sciences (MAT) | 26.9 | 39.6 | 6.5 |
| Chemistry (CHE) | 25.2 | 26.4 | 1.3 |
| Agricultural and veterinary sciences (AVS) | 20.1 | 43.8 | 3.2 |
| Medicine (MED) | 18.6 | 41.7 | 2.9 |
| Civil engineering (CEN) | 15.3 | 59.3 | 2.8 |
| Industrial and information engineering (IIE) | 13.4 | 50.0 | 1.5 |
| Total | 23.8 | 38.1 | 4.1 |

The low value of propensity registered in the engineering disciplines, where over half of academics did not collaborate at the international level in this period, can be explained in light of the previously-noted prevalence of intramural collaborations. However, this situation could also reflect a circumstance unique to the disciplines in the Italian context, seeing as Tijssen and Van Wijk (1998) have shown that in the ICT engineering specialties, Italy has a percentage of domestic publication that is higher than for other European nations. The maximum values of propensity to foreign collaboration, as confirmed by the Mann-Whitney U test results, are observed in Physics and in Earth sciences. This result is amply confirmed by numerous studies in the literature (Luukkonen et al., 1992; Glanzel and Schubert, 2005; Olmeda-Gómez et al., 2006; Abramo et al., 2011) and is essentially due to the fact that much research in these disciplines requires observations that are so complex and equipment that is so costly that can only be obtained through international collaborations. The differences between different UDAs in propensity for international collaboration can also be explained on the basis of the classification of basic and applied disciplines, suggested by Frame and Carpenter (1979). This classification is not based only on the equipment used, but also the research themes, which for basic disciplines would have a broader and more general horizon, less linked to specific national contexts. This would explain the high propensity to international collaboration registered in Mathematics and computer sciences, as seen in our study and in those of Archibugi and Coco (2004),

Wagner (2005) and Abramo et al. (2011). In fact the research conducted in this discipline generally does not require major equipment, but is typically of supranational interest. On the other hand, while Medicine shows a high propensity to extramural domestic collaboration, which is again confirmed by other research (Olmeda-Gómez et al., 2006; Thijs and Glänzel, 2010), it shows a relatively low percentage of collaboration at the international level. This is due to great differences between individual national health systems, which heavily influence research policies in the medical field (Hoekman et al., 2010). Considering the entire Italian university system, the average propensity to collaborate with foreign organizations is notably lower than the propensity to intramural or extramural domestic collaboration. This result is in line with those obtained by Gazni et al. (2012), from over 13,000,000 world-wide publications over the 2000-2009 period, and by Haiqi and Hong (1997) from over 100,000 publications by Chinese scientists in 1993. On the other hand, studies by Gazni and Didegah (2011), on over 120,000 publications for the 2000-2009 period by scientists at the University of Harvard, and by Olmeda-Gómez et al. (2006), on the entire scientific production of the Spanish university system from 2000 to 2004, show a higher percentage of international publications compared to the Italian context.

On the basis of the results obtained we can classify the analyzed UDAs into four clusters characterized by specific propensities to collaborate:

- a cluster characterized by the lowest general propensities to collaborate and a low value of propensity for intramural collaboration (UDAs of Mathematics and computer sciences and Economics and statistics);
- a cluster characterized by high propensity for intramural collaboration and the minimum values of propensity for extramural collaboration, both at domestic and international level (Civil engineering and Industrial and information engineering);
- a cluster characterized by the maximum values of general propensity to collaborate, with the collaborations realized principally at the national level (Chemistry and Medicine, with possible association of Agricultural and veterinary sciences and Biology);
- a cluster characterized by a low value of propensity to collaborate at the intramural level and a high propensity for extramural collaboration, especially at the international level (Physics, Earth sciences, with possible association of Pedagogy and psychology).

*4.2 The propensity to collaborate in the different forms for the various fields of Physics*

In the preceding section we have shown how the propensity to collaborate varies notably between the various disciplines, to the extent that they can be classified in clusters that are very different from each other. However our approach can also be applied within the individual disciplines, to reveal if there are significant differences between the various fields that compose them. For illustrative purposes, we will carry out the analysis for the eight SDSs of Physics.

Table 6 presents the structural characteristics for the Physics SDSs, in the same way that Table 1 presented this data for the various disciplines. From the results in this table and even more so from the results in Table 7, we see how the propensity to collaborate is not homogenous between the various SDSs. Concerning propensity for intramural collaboration, some SDSs (Theoretical physics, Mathematical models and methods, and Astronomy and astrophisycs) are characterized by a very low value, signaling how the academics that belong to these SDSs tend to collaborate predominantly at the extramural level, perhaps in part for the lack of a sufficient number of colleagues within their own universities who work on similar themes. In the extramural collaborations, there is distinction between two classes of SDSs: i) those of a more theoretical nature that involve less costly equipment (FIS/02, FIS/03 and especially FIS/06, FIS/07, FIS/08), characterized by a lower extramural propensity, especially at the international level; ii) those more justifiably considered "big science" (FIS/01, FIS/04 and FIS/05), where the propensity for international collaboration reaches very high levels.

*Table 6: Principle characteristics of the population of academics in the SDSs of Physics*

| SDS | Publications | Research staff | | |
| --- | --- | --- | --- | --- |
| | | Total | Productive | Collaborative |
| Experimental Physics (FIS/01) | 9,818 | 1,103 | 1,001 (90.8%) | 995 (90.2%) |
| Physics of Matter (FIS/03) | 6,736 | 502 | 475 (94.6%) | 473 (94.2%) |
| Theoretical Physics, Mathematical Models and Methods (FIS/02) | 3,468 | 394 | 348 (88.3%) | 335 (85.0%) |
| Astronomy and Astrophysics (FIS/05) | 3,333 | 209 | 188 (90.0%) | 188 (90.0%) |
| Applied Physics: Cultural Heritage, Environment, Biology and Medicine (FIS/07) | 3,091 | 371 | 332 (89.5%) | 331 (89.2%) |
| Nuclear and Subnuclear Physics (FIS/04) | 2,175 | 181 | 170 (93.9%) | 169 (93.4%) |
| Physics for Earth and Atmospheric Sciences (FIS/06) | 433 | 65 | 60 (92.3%) | 60 (92.3%) |
| Didactics and History of Physics (FIS/08) | 145 | 48 | 28 (58.3%) | 24 (50.0%) |
| Total | 23,702* | 2,873 | 2,602 (90.6%) | 2,575 (89.6%) |

*The total is less than the sum of those from the different SDSs, since 5,497 publications are in co-authorship by academics from different SDSs.

*Table 7: Mean propensity of academics to collaborate in different forms, per SDS of Physics (percentage values)*

| SDS | C | CI | CED | CEF |
| --- | --- | --- | --- | --- |
| Applied Physics: Cultural Heritage, Environment, Biology and Medicine (FIS/07) | 98.7 | 77.3 | 66.4 | 33.4 |
| Physics for Earth and Atmospheric Sciences (FIS/06) | 98.1 | 60.4 | 60.2 | 36.5 |
| Experimental Physics (FIS/01) | 98.0 | 76.7 | 78.6 | 60.1 |
| Physics of Matter (FIS/03) | 97.3 | 65.8 | 64.1 | 41.8 |
| Astronomy and Astrophysics (FIS/05) | 96.7 | 47.5 | 67.6 | 64.3 |
| Nuclear and Subnuclear Physics (FIS/04) | 96.3 | 70.9 | 87.2 | 70.4 |
| Theoretical Physics, Mathematical Models and Methods (FIS/02) | 90.7 | 39.2 | 72.5 | 47.3 |
| Didactics and History of Physics (FIS/08) | 81.3 | 59.8 | 43.4 | 26.8 |
| Total | 96.6 | 66.7 | 72.5 | 51.7 |

*4.3 Correlation between propensities to collaborate in different forms*

The results concerning propensities to collaborate in different forms show that in some UDAs (Section 4.1), academics tend to collaborate both with scientists from the same university and from other organizations. This tendency is especially seen in the fields of research that require either a critical mass of internal resources at the individual university or an adequate network of external partners. In other cases, the different forms of coordination required under each form of collaboration could actually lead scientists to favor one form of collaboration at the expense of another. Further, in some disciplines, the tendency to produce publications with a limited number of authors could further lead to the choice of a single form of collaboration. To evaluate the relations between the propensities to collaborate in different forms, we use the R rcorrr function (R Development Core Team, 2012) to calculate the Spearman non-parametric correlation between the values obtained from each academic for the four indicators *C*, *CI*, *CED* and *CEF*. The results, presented in Table 8, permit evaluation both at the general level and for each UDA if and how the four forms of collaboration are correlated among each other.

*Table 8: Spearman correlation between the indicators of propensity to collaborate, per UDA*

| UDA | C-CI | C-CED | C-CEF | CI-CED | CI-CEF | CED-CEF |
|---|---|---|---|---|---|---|
| AVS | 0.20*** | 0.10*** | 0.05** | -0.35*** | -0.25*** | 0.00 |
| BIO | 0.23*** | 0.12*** | 0.02 | -0.21*** | -0.25*** | -0.03 |
| CEN | 0.43*** | 0.14*** | 0.10*** | -0.44*** | -0.25*** | 0.04 |
| CHE | 0.25*** | 0.12*** | 0.04* | -0.28*** | -0.21*** | -0.01 |
| EAR | 0.21*** | 0.21*** | 0.11*** | -0.28*** | -0.25*** | 0.08** |
| ECS | 0.41*** | 0.35*** | 0.27*** | -0.20*** | -0.23*** | -0.03 |
| IIE | 0.38*** | 0.12*** | 0.06*** | -0.44*** | -0.32*** | 0.15*** |
| MAT | 0.43*** | 0.25*** | 0.20*** | -0.32*** | -0.30*** | 0.03 |
| MED | 0.17*** | 0.10*** | 0.02 | -0.29*** | -0.26*** | 0.07*** |
| PHY | 0.33*** | 0.34*** | 0.20*** | -0.02 | -0.04 | 0.37*** |
| PPS | 0.24*** | 0.21*** | 0.14*** | -0.30*** | -0.30*** | -0.03 |
| Total | 0.36*** | 0.21*** | 0.08*** | -0.21*** | -0.27*** | 0.12*** |

*Significance level: \*\*\* p < 0.001; \*\* p < 0.01; \* p < 0.05*

The data show how the propensity to collaborate *C* is positively correlated with all other indicators. This result is common to all the UDAs, although with different intensities, and suggests how the scientists who adopt these forms of collaboration to a greater extent also register a greater tendency to collaborate in general. The link between propensity to collaborate *C* and propensity to collaborate at the international level *CEF* is more complex to interpret, because the variability among UDAs is very high. In particular, in Economics and statistics the correlation is strongly significant and positive, since the academics who collaborate – seen in Table 4 to be only 60% in this discipline – tend to activate ties with foreign organizations.

Analyzing the links between the three indicators of propensity to collaborate in specific forms *CI*, *CED* e *CEF*, we note how intramural collaboration seem negatively correlated to extramural collaborations. This result is encountered in all UDAs and indicates how intramural collaborations are seen as an alternative and not a complement to extramural collaboration. This tendency is less strong only in Physics, where we observe a strong significant and positive correlation between propensity to extramural collaboration at the national and international levels. This correlation between *CED* and *CEF* is significant and positive in four UDAs only. In substance, particularly in certain disciplines, extramural collaborations are often extended to the international level, probably in an attempt to construct an adequate network of partners for development of the research.

**5. Main differences between propensity to collaborate, observed for individual scientists, and incidence of collaboration based on publications**

The methodological innovation in the current work concerns the use of the single scientist as the unit of base analysis for the evaluation of the different means of collaboration. In the literature, the phenomenon of collaboration is generally analyzed through indicators of "incidence", that measure the ratio of the number of publications characterized by specific features – for example, that of being the result of international collaborations – to the total of publications referred to the context of analysis. The analyses of collaborations across disciplines thus refers to the classification of publications on the basis of the journal subject category. For example, the analysis of international collaborations by researchers in a given country (for example Italy) working in a specific discipline (for example Physics) is based on indicators of incidence, where an example of a particular indicator would be the ratio of the number of publications in Physics-subject-category journals having an address list that includes at least one "Italy" and "non-Italy" country, to the total of Italian publications in the same Physics journals.

A comparative analysis between this approach and the one used in the current work is not readily apparent, given that the classification system for the publications (subject categories) and that for the Italian academic authors (SDSs) do not fully correspond. However it is possible to associate each publication with one or more SDSs (and then their UDAs) on the basis of those of the authors, use these associations to carry out the measurement of the "incidence" indicators, and then proceed to a comparison with the indicators of propensity. In this section we adopt the procedure described, to calculate the indicators of incidence in each UDA for the four forms of collaboration studied in this work.

These indicators permit the evaluation of the extent of the form of collaboration in any given context, but unlike the "homologous" indicators of propensity do not provide exact information on the behavior of the individual scientists, meaning the subjects that determine the phenomenon. The indicators of propensity and the matching indicators of incidence are not only different from a conceptual point of view but can also register extremely different values, whenever either the productivity or collaboration intensity (for example relative to propensity to collaborate with foreign colleagues) are distributed in a non-homogenous manner in the population of interest – situations which are completely normal. Concerning the first factor, previous measurement conducted by the authors, for the Italian academic system over the 2004-2008 period, showed that 29% of researchers produced 71% of total national scientific output (Abramo et al., 2011a). This means that potential analyses of collaborative patterns based on counting of publications would depend strongly (71% of total observations) on the behavior of small share of individuals (29% of total researchers). Instead with our approach, each researcher has the same weight in determining the average value of propensity to collaborate in a field. Taking the example of an analysis of collaborative patterns where the objective is to establish benchmarks for evaluation of individual scientist behavior, the use of "incidence" indicators based on counting of publications would lead to distorted results, which would in any case be less reliable than those guaranteed by the indicators of propensity, proposed in the present work.

We now quantify the shifts between measurements conducted under the two approaches. For ease of comprehension, let us first consider the case of a population made of two researchers only, α and β, belonging to the same UDA. Their output (Table 9) is: 23 publications (*p*) for α, of which 4 co-authored by scientists belonging to foreign organizations (*cefp*), and 13 publications for β, of which 3 co-authored by scientists belonging to foreign organizations. Furthermore, 8 of their publications present both α and β in the byline, of which 3 co-authored by scientists belonging to foreign organizations. The propensity to extramural collaboration at the international level (*CEF*) of α and β is, respectively, of 17.4% and 23.1%, which leads to an average of 20.2%. To calculate the incidence instead, one divides the UDA's publications co-authored by scientists belonging to foreign organizations (4) by the total UDA's publications (28), which results in an incidence of 14.3%.

*Table 9: Example of the estimation of the propensity and incidence indicators of the extramural collaboration at the international level*

|      | Researcher α | Researcher β | UDA |
|------|--------------|--------------|------|
| p    | 23           | 13           | 28   |
| cefp | 4            | 3            | 4    |
| CEF  | 17.4%        | 23.1%        | 14.3% |

Repeating the procedure for all scientists of the same UDA, for all UDAs, we obtain data shown in Table 10, which reports differences (*Δ*) between the indicators of average propensity *C*, *CI*, *CED* and *CEF* and the matching indicators of incidence.

*Table 10: Difference between mean values of propensity to collaborate and "homologous" indicators of incidence (percentage values)*

| UDA   | C    | Δ   | CI   | Δ    | CED  | Δ   | CEF  | Δ    |
|-------|------|-----|------|------|------|-----|------|------|
| MED   | 99.4 | 0.8 | 81.1 | 15.4 | 62.3 | 2.9 | 18.6 | -9.5 |
| CHE   | 99.2 | 0.7 | 83.5 | 12.2 | 50.1 | 4.4 | 25.2 | -11.2|
| AVS   | 99.1 | 0.4 | 81.2 | 11.2 | 47.1 | 0.9 | 20.1 | -8.6 |
| BIO   | 99.1 | 0.8 | 78.8 | 11.6 | 57.7 | 1.3 | 27.1 | -6.5 |
| EAR   | 97.6 | 0.5 | 62.0 | 12.6 | 58.0 | 1.6 | 32.8 | -8.8 |
| IIE   | 97.1 | 0.8 | 82.2 | 5.6  | 24.5 | 1.3 | 13.4 | -5.0 |
| PPS   | 96.7 | 0.0 | 59.6 | 11.4 | 51.5 | 1.7 | 31.8 | -9.5 |
| PHY   | 96.6 | 0.9 | 66.7 | 16.1 | 72.5 | 6.1 | 51.7 | 1.4  |
| CEN   | 94.3 | 1.7 | 73.4 | 7.0  | 25.8 | 2.1 | 15.3 | -7.2 |
| MAT   | 89.1 | 1.2 | 54.1 | 2.8  | 33.6 | 5.2 | 26.9 | -4.9 |
| ECS   | 84.0 | 0.3 | 44.0 | 2.5  | 37.8 | 0.0 | 27.5 | -6.5 |
| Total | 97.2 | 1.0 | 75.4 | 13.3 | 50.3 | 5.6 | 23.8 | -7.8 |

At the general level, the indicators of propensity and homologous indicators of incidence result as strongly correlated. However, for specific forms of collaboration we observe shifts that reach substantial levels. For intramural collaboration, we see that the indicator of incidence notably underestimates the average propensity registered for the academics. The difference between the two indicators is an average of 13.3 points and is more than 10 points in seven of the 11 UDAs considered. In contrast, in the analysis of extramural collaborations at international level, the indicator of incidence notably overestimates the average propensity registered by the academics: the difference between the two indicators averages 7.8 points, with a maximum of 11.2 points in the Chemistry UDA.

## 6. Conclusions

The analysis of the various forms of research collaboration has attracted attention from a number of scholars, interested in determining if and how the patterns of collaboration might vary across different fields, as well as in formulating hypotheses on the factors that could determine such patterns. Until now this has been done by resorting to indicators of incidence, based on counting of publications. In this work, the authors instead propose an innovative methodological approach based on the use of the single scientist as base unit of observation. This approach offers several advantages, first of all permitting large scale analysis of inter-disciplinary collaborations.

More generally, the proposed approach certainly permits a more truthful picture of the propensity of researchers to collaborate in the various forms, either with their direct colleagues or with other organizations: in fact, basing the quantification of the collaboration phenomenon on counting of publications implies obvious distortions in the case where productivity, apart from collaboration intensity, is not distributed in homogenous fashion (the real-life situation) among researchers within the various fields that are analyzed.

The implementation of reliable collaboration measurement systems is in fact fundamental for correct ex-ante definition and ex-post control of policies to develop, modify or maintain the conditions for different forms of collaboration within any reference context. It is clear that many nations have policies intended to foster collaboration among scientists, given the positive returns that collaboration can ensure in capacities to produce and diffuse new knowledge. The measure of propensity to collaborate for the individual scientist permits verifying the effect of such policy on the actors that are the ultimate target of the policy. Further, beginning from the data on individual scientists, it is possible to obtain the measurement of the propensity to collaborate for the individual's research group and organizational unit at increasing levels, which in turn can be the object of specific policy.

In summary, our approach supports the implementation of policy directed at influencing scientific collaborations, in a more appropriate manner than the measures proposed in the literature until now. The application of our approach to Italian academics' research activity has permitted measurement of their propensity to collaborate using different forms, in the various fields. The results can be used to both measure the effects of the policies adopted in the past by single universities, or the entire research system, and to formulate new policies aimed at fostering collaborations, taking into account the intrinsic characteristics of each discipline.

The results obtained with the method we proposed also permit the individual scientists to self-evaluate their own propensities to collaborate in different forms, comparing with the average propensity of their colleagues that belong to the same field, and also to analyze their situation relative to any available incentives (or disincentives) for collaboration.

The proposed methodological approach lends itself to further development at various levels. For example, it is clearly of interest to investigate the determinants of the different propensities to collaborate for scientists in the same field. Specific questions are whether gender or academic rank can be causes of different collaboration behavior, or if geographic location can influence the propensity or forms of collaboration. The authors will examine these more detailed questions in further research.

*Scientometrics*, 62(1), 3–26

Wagner, C.S., Leydesdorff, L. (2005). Network structure, self-organization, and the growth of international collaboration in science. *Research Policy*, 34(10), 1608–1618.

Yoshikane, F., Kageura, K. (2004). Comparative analysis of coauthorship networks of different domains: The growth and change of networks. *Scientometrics*, 60(3), 433–444.

Zitt, M., Bassecoulard, E., Okubo, Y. (2000). Shadows of the past in inter-national cooperation: Collaboration profiles of the top five producers of science. *Scientometrics*, 47(3), 627–657.
23